\documentclass[12pt]{iopart}

\usepackage{graphicx,epstopdf,color}
\begin{document}


\title{Non-linear spectroscopy of rubidium: An undergraduate experiment }

\author{V Jacques, B Hingant, A Allafort, M Pigeard and J F Roch}

\address{D\'epartement de Physique, Ecole Normale Sup\'erieure de Cachan\\
 61 avenue du Pr\'esident Wilson, 94235 Cachan, France.}
\date{\today}
\ead{vincent.jacques@lpqm.ens-cachan.fr}

\begin{abstract}
In this paper, we describe two complementary non-linear spectroscopy methods which both allow to achieve Doppler-free spectra of atomic gases. First, saturated absorption spectroscopy is used to investigate the structure of the $5{\rm S}_{1/2}\rightarrow 5{\rm P}_{3/2}$ transition in rubidium. Using a slightly modified experimental setup, Doppler-free two-photon absorption spectroscopy is then performed on the $5{\rm S}_{1/2}\rightarrow 5{\rm D}_{5/2}$ transition in rubidium, leading to accurate measurements of the hyperfine structure of the $5{\rm D}_{5/2}$ energy level. In addition, electric dipole selection rules of the two-photon transition are investigated, first by modifying the polarization of the excitation laser, and then by measuring two-photon absorption spectra when a magnetic field is applied close to the rubidium vapor. All experiments are performed with the same grating-feedback laser diode, providing an opportunity to compare different high resolution spectroscopy methods using a single experimental setup. Such experiments may acquaint students with quantum mechanics selection rules, atomic spectra and Zeeman effect.
\end{abstract}


\maketitle
\section{Introduction}

The development of laser diode devices about thirty years ago is nowadays regarded as an indisputable breakthrough either for the growth of a broad range of technological applications or for its huge impact on atom physics research, from high resolution spectroscopy to laser cooling techniques. Although the spectral properties and the tunability of these lasers may prove unsuited for some applications, an efficient device can be built by stabilizing the diode using an optical feedback from an external grating operating in Littrow configuration~\cite{GratHansch,Arnaold}. One can thus easily obtain a low-cost narrow band laser with several tens of milliwatts of optical power and a mode-hop free frequency tuning which can be larger than $50$~GHz.\\
\indent Such tunable laser diodes are useful tools for teaching atom physics in advanced undergraduate laboratory courses. Indeed, many experiments have been developed for undergraduates using these devices in the past years, ranging from atomic hyperfine structure studies of Rubidium and Cesium~\cite{MacAdam,Preston}, to interferometric measurements of the resonant absorption and refractive index in rubidium gas~\cite{Libbrecht}, temperature dependance of Doppler-broadening~\cite{Leahy}, or observation of Faraday effect~\cite{VanBaak} and two-photon spectroscopy in Rubidium~\cite{Olson}.\\
\indent In this paper, we use a simple grating-feedback laser diode to investigate the hyperfine structure of the $5{\rm P}_{3/2}$ and $5{\rm D}_{5/2}$ excited states in rubidium. This is achieved following two different methods of Doppler-free high resolution spectroscopy. First, saturated absorption spectroscopy is performed on the $5{\rm S}_{1/2}\rightarrow 5{\rm P}_{3/2}$ transition in rubidium to measure the hyperfine structure of the $5{\rm P}_{3/2}$ excited-state. Using the same laser diode and a slightly modified experimental setup, Doppler-free two photon absorption spectroscopy is then performed on the transition $5{\rm S}_{1/2}\rightarrow 5{\rm D}_{5/2}$, leading to accurate measurements of the hyperfine structure of the $5{\rm D}_{5/2}$ energy level. These two techniques, both based on non-linear interaction of light with atoms, are complementary since they enable to probe atomic transitions following different electric dipole selection rules. Indeed, for single-photon transitions involved in saturated absorption spectroscopy, the orbital angular momentum $l$ must satisfy the selection rule $\Delta l=\pm 1$. Consequently, transitions have to involve ground and excited states with opposite parity. Contrarily, selection rules for two-photon transitions become $\Delta l=0, \pm 2$, allowing to investigate transitions between levels of identical parity.\\
\indent In the following, we begin by providing a general description of the experimental apparatus used to perform the experiments. We then briefly discuss Doppler broadening and describe Doppler-free saturation absorption spectroscopy of the $5{\rm S}_{1/2}\rightarrow 5{\rm P}_{3/2}$ transition in rubidium, which is the simplest undergraduate experiment using tunable laser diode~\cite{MacAdam,Preston}. Afterwards, the hyperfine structure of the excited state $5{\rm D}_{5/2}$ is probed using Doppler-free two-photon absorption spectroscopy. In addition, electric dipole selection rules of the two-photon transition are investigated, first by modifying the polarization of the excitation laser, and then by measuring two-photon absorption spectra when the rubidium vapor cell si placed in a magnetic field. Such experiments provide an opportunity to compare different high resolution spectroscopy methods using a single experimental setup and may acquaint students with quantum mechanics selection rules, atomic spectra and Zeeman effect.

\section{Experimental apparatus}

\begin{figure}  [t]
\centerline{\includegraphics[width=10cm]{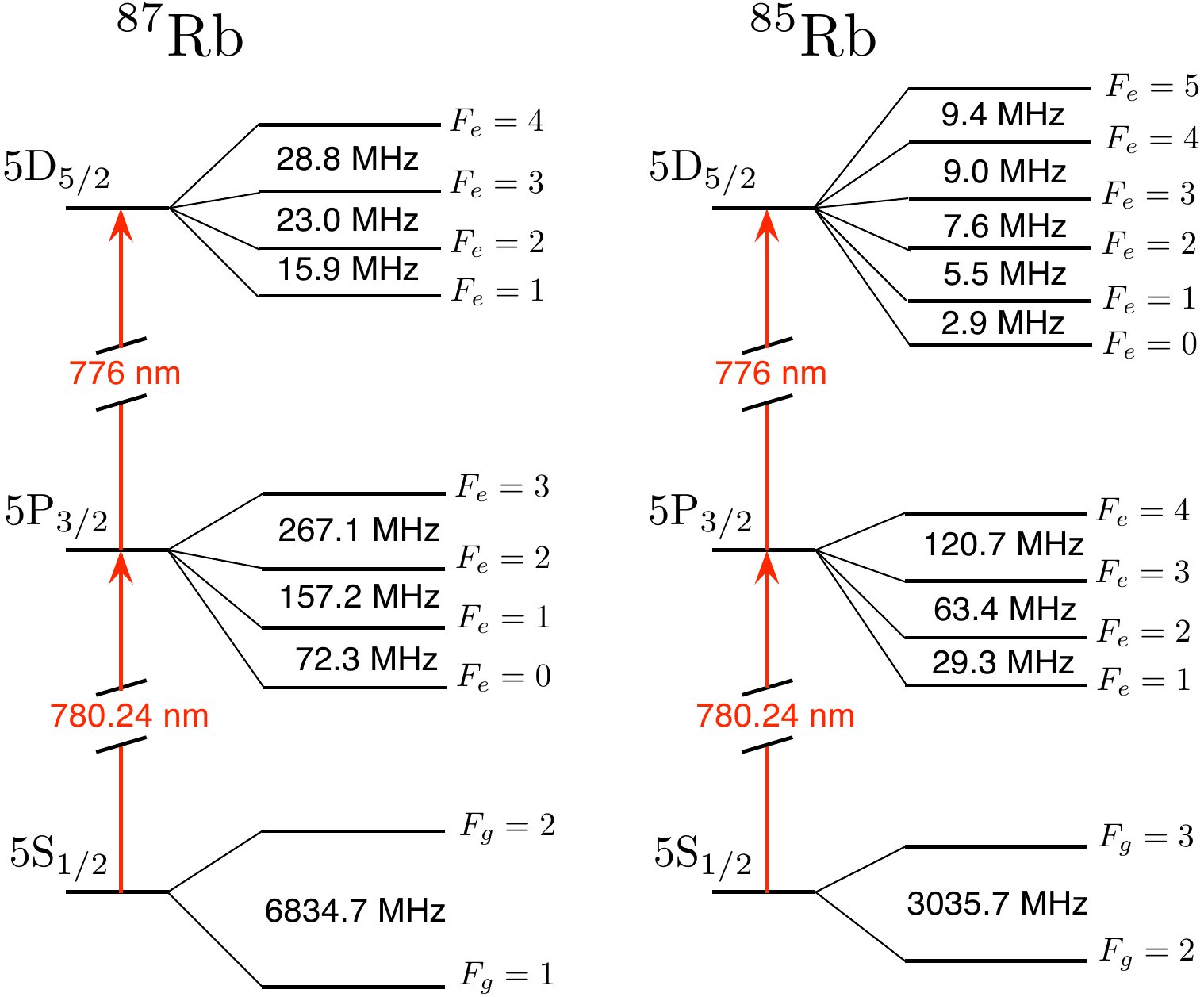}}
\caption{Energy level diagram for $^{85}\rm Rb$ and $^{87}\rm Rb$.}
\label{Rb}
\end{figure}

\indent All experiments are performed with a commercial grating-feedback laser diode (Toptica Photonics, DL100) operating in Littrow configuration~\cite{Conroy}. Such laser diode, stabilized in temperature, has a typical free-running wavelength of $780$ nm with $1$ MHz linewidth and an output power of $30$ mW. Besides, a $30$ dB optical isolator (Electro-Optics Technologies) is used to prevent from laser instability caused by optical feedback in the laser diode cavity.\\
\indent Preliminary tuning of the laser diode emission wavelength close to a resonant transition of rubidium is achieved by manually changing the length of the external cavity and by monitoring the output wavelength using a commercial wavelength-meter~\cite{Fox}. Scanning the frequency of the laser diode is then realized by applying a voltage ramp to a piezoelectric transducer which changes the external cavity length and thus the emission wavelength. With Toptica DL100 commercial diode, a mode-hop free frequency tuning around $20$~GHz is easily achieved.\\
\indent Part of laser emission is constantly sent into a homemade Fabry-Perot Interferometer (FPI), with a $750$ MHz free-spectral-range and a finesse $\mathcal{F}=10$, which transmission peaks provide a frequency reference for the measurements of rubidium spectral features. A rubidium vapor cell containing $^{85}\rm Rb$ ($72\%$ natural abundance) and $^{87}\rm Rb$ ($28\%$ natural abundance) from Thorlabs is used. The energy level diagram of the relevant transitions in rubidium is shown in figure~\ref{Rb}.

\section{Doppler broadening of absorption spectra}
\label{sectionDop}

\indent In conventional laser spectroscopy, atomic hyperfine structure is often hidden by inhomogeneous Doppler broadening. Indeed, when atoms in a vapor cell are irradiated by a laser beam at frequency $\nu_{L}$ in the laboratory frame of reference, they experiment in their own frame a Doppler shifted laser frequency $\nu$, related to the atom velocity ${\rm v}_{z}$ along the incident light direction $z$ as
\begin{equation}
\label{dop1}
\nu=\nu_{L}\left[1-\frac{{\rm v}_{z}}{c}\right] \ ,
\end{equation}
where $c$ is the speed of light. This formula is valid in the approximation of non-relativistic atoms ${\rm v}_{z}\ll c$.\\
\indent As a result, by scanning the laser frequency $\nu_{L}$ around an atomic transition at frequency $\nu_{0}$, the class of atoms with velocity ${\rm v}_{z}$ absorbs light when the condition
\begin{equation}
\label{dop2}
\nu_{L}=\frac{\nu_{0}}{1-\frac{{\rm v}_{z}}{c}} \ 
\end{equation}
is fulfilled. For atoms in a vapor cell, the probability distribution of velocities $p({\rm v}_{z})$ follows a Maxwell-Boltzmann distribution 
\begin{equation}
\label{MaxBol}
p({\rm v}_{z})=\sqrt{\frac{m}{2\pi k_{b}T}} \exp(\frac{m{\rm v}_{z}^{2}}{2k_{b}T}) \ ,
\end{equation}
where $k_{b}$ the Boltzmann constant, $T$ the absolute temperature and $m$ the mass of the atoms. By substituting equation~(\ref{dop2}) into equation~(\ref{MaxBol}), the relative number of atoms $N$ which are resonant with the laser at frequency $\nu_{L}$, is given by the Gaussian function 
\begin{equation}
N=\exp\left[\frac{mc^{2}}{2k_{b}T}\left(\frac{\nu_{0}-\nu_{L}}{\nu_{L}}\right)^{2}\right] \ .
\end{equation}
\indent This distribution is directly translated in a Gaussian shape of the atomic medium absorption profile, centered at the resonant frequency $\nu_{0}$ and with a full-width at half-maximum (FWHM) $\Delta\nu_{\rm dop}$ given by 
\begin{equation}
\label{gaussien}
\Delta\nu_{\rm dop}=2\nu_{0}\sqrt{\frac{2k_{B}T\ln 2}{mc^{2}}} \ .
\end{equation}
\indent For rubidium atoms, $\Delta\nu_{\rm dop} \approx 500$ MHz at room temperature, which is smaller than the energy splitting between ground state $5S_{1/2}$ hyperfine levels but much bigger than the one of the excited state $5P_{3/2}$ (See Fig.~\ref{Rb}).\\
 \begin{figure}  [t]
\centerline{\includegraphics[width=10cm]{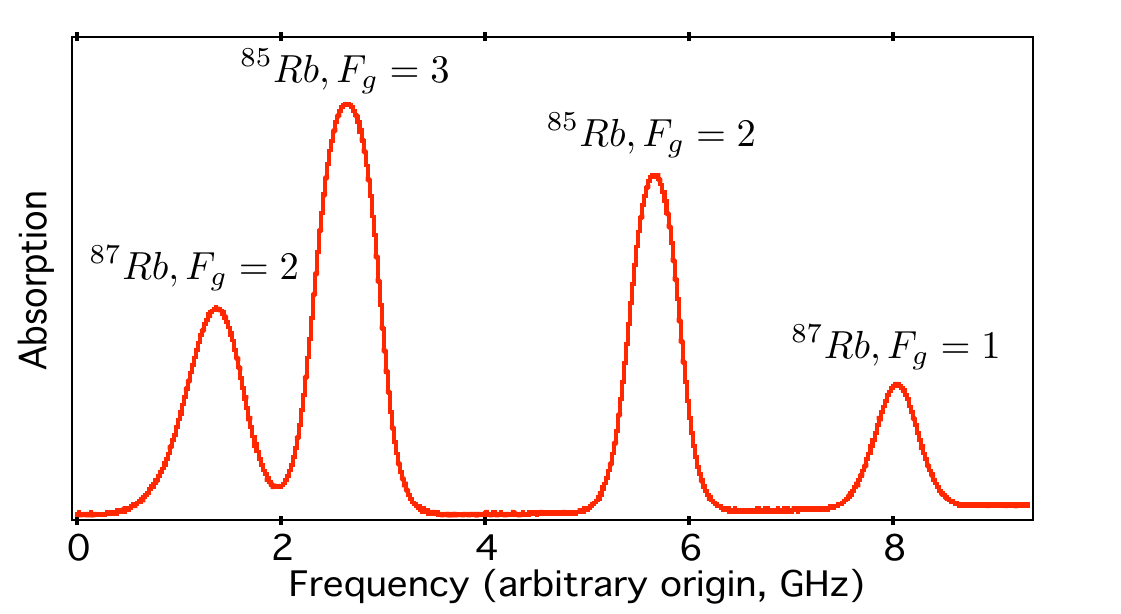}}
\caption{Doppler-broadened absorption spectrum related to the transition $5{\rm S}_{1/2}\rightarrow 5{\rm P}_{3/2}$ in rubidium at room temperature. The ground-state hyperfine splitting is measured to be $3.02\pm 0.03$ GHz for $^{85}\rm Rb$ and $6.80\pm 0.07$ GHz for $^{87}\rm Rb$. Excited-state hyperfine structure is hidden by Doppler broadening.}
\label{Doppler}
\end{figure}
\indent In order to experimentally observe Doppler broadened absorption spectra, the extended-cavity diode laser is tuned close to the rubidium transition $5{\rm S}_{1/2}\rightarrow 5{\rm P}_{3/2}$ at the wavelength $\lambda=780.24$ nm (vacuum wavelength) and directed through the rubidium vapor cell. A typical absorption profile for the $5{\rm S}_{1/2}\rightarrow 5{\rm P}_{3/2}$ transition in rubidium is depicted in figure~\ref{Doppler}. Ground state hyperfine levels are resolved whereas the hyperfine structure of the $5P_{3/2}$ excited-state remains hidden by Doppler broadening. Gaussian fit of each absorption line gives a FWHM of $518\pm 15$ MHz as expected for Doppler broadening at room temperature~\cite{Dop}.

\section{Doppler-free saturated absorption spectroscopy}
\label{SatRes}
\subsection{Principle}

\indent In the early seventies, Theodor W. H$\ddot{{\rm a}}$nsch and Christian Bord\'e independently introduced a method using non-linear interaction of laser light with atoms to achieve Doppler-free spectra of atomic gases~\cite{Hansch_PRL71,Hansch_PRL71bis,Borde_70}. This technique, commonly called saturated-absorption spectroscopy, has revolutionized spectroscopy studies~\cite{Schalow_Nobel} and is now widely used in atom physics research as a versatile method to lock laser frequencies on atomic transitions. \\
\indent In such a technique, two counterpropagating laser beams at identical frequencies $\nu_{L}$ are interacting with atoms in a vapor cell. One beam is noted $P$ as pump beam, and the other one, weaker, is noted $S$ and usually called probe beam. When the laser frequency $\nu_{L}$ is different from the frequency $\nu_{0}$ of the atomic transition ({\it e.g.} $\nu_{L}>\nu_{0}$), the pump beam $P$ interacts with a group of atoms with velocity ${\rm v}_{z}$, whereas the counterpropagating probe beam $S$ excites the symetric group of velocity $-{\rm v}_{z}$. As a result, the two laser beams interact with different classes of atom and the absorption spectra are identical to the one obtained using a single laser excitation scheme. However, under the particular circumstance where $\nu_{L}=\nu_{0}$, both beams interact with the same class of atoms of velocity ${\rm v}_{z}=0$. In that case, the pump beam $P$ saturates the atomic transition, depleting the number of atom with velocity ${\rm v}_{z}=0$ in the ground state. As a result, the absorption of the counterpropagating probe beam is decreased, as most part of the atoms are already excited by the pump beam. The absorption of the probe beam finally presents a Doppler broadened profile, on which is superimposed a dip corresponding to the resonant frequency $\nu_{0}$, for which both beams are interacting with the same class of atoms ${\rm v}_{z}=0$.\\
\indent Using Bloch equations theory in the approximation of a two level atom, the dip is shown to be Lorentzian with a linewidth $\Delta \nu$ (FWHM) given by the relation~\cite{Cahuzac}
\begin{equation}
\label{Bloch}
\Delta \nu=2\sqrt{\gamma^{2}+\frac{\Omega_{1}^{2}\gamma}{\Gamma_{\rm sp}}} \ ,
\end{equation}
where $2\gamma$ is the natural linewidth of the transition, $\Gamma_{\rm sp}$ the spontaneous emission rate and $\Omega_{1}$ the Rabi frequency associated with the pump beam $P$ which saturates the atomic transition. In principle, such method should then allow to reach an asymptotic value of the natural linewidth of the transition by decreasing the saturating pump beam intensity.

\subsection{Crossover resonance}

\indent We now consider two atomic transitions at frequencies $\nu_{1}$ and $\nu_{2}$ ($\nu_{1}<\nu_{2}$) involving a common ground state, with a frequency difference smaller than the Doppler broadening $\Delta\nu_{\rm Dop}$.\\
\indent When the laser frequency is set exactly midway between the two resonances, $\nu_{L}=(\nu_{1}+\nu_{2})/2$, the pump beam interacts with two classes of atoms of opposite velocities ${\rm v}_{z}$ and $-{\rm v}_{z}$. Positive velocity atoms ${\rm v}_{z}$ experiment the pump beam redshifted to the lower transition frequency $\nu_{1}=\nu_{L}(1-{\rm v}_{z}/c)$ whereas atoms with negative velocity $-{\rm v}_{z}$ see the pump beam blueshifted to the higher resonance frequency $\nu_{2}=\nu_{L}(1+{\rm v}_{z}/c)$. This results in a depletion of the number of atom with velocities $\pm {\rm v}_{z}$ in the ground state. At the same frequency, the probe beam interacts with exactly the same group of atoms, but in opposite way as atoms with velocity $-{\rm v}_{z}$ (resp. ${\rm v}_{z}$) are resonant with the transition $\nu_{1}$ (resp. $\nu_{2}$). Then, the probe beam absorption profile once again shows a dip, more intense than the one observed for standard resonance as it involves two groups of atom velocities. This dip, usually called a crossover resonance, is peculiar to saturated absorption spectroscopy.

\subsection{Experimental setup and results}

 \begin{figure}  [t]
\centerline{\includegraphics[width=10cm]{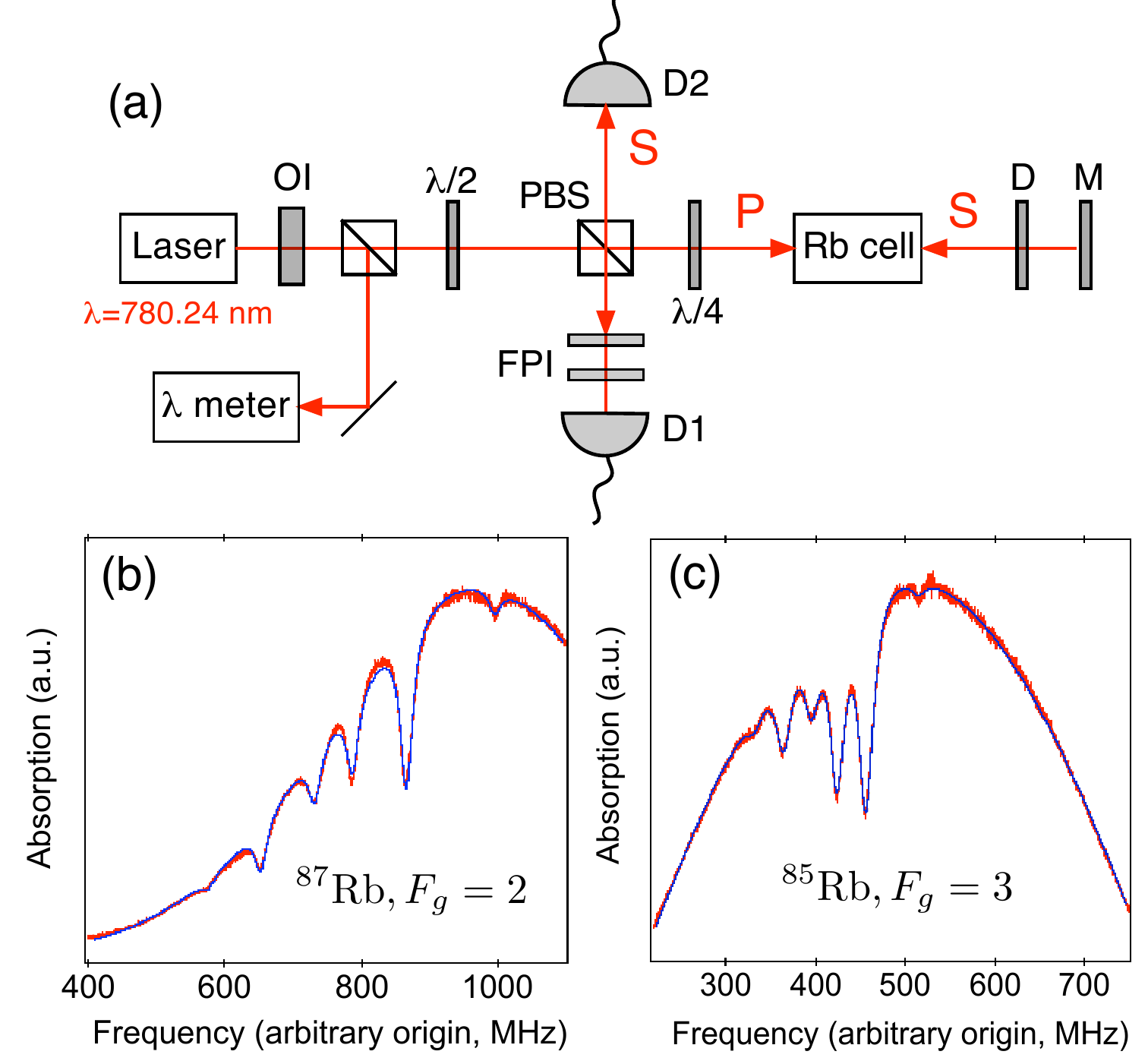}}
\caption{Doppler-free saturated absorption spectroscopy of rubidium. {\bf (a)-} Experimental setup. OI: optical isolator, $\lambda /2$: half-wave plate, $\lambda /4$: quarter-wave plate, PBS: polarization beamsplitter, FPI: Fabry-Perot Interferometer, D: optical density ($1\%$), M: mirror, D1 and D2: photodiodes. P and S denote the pump and probe beam respectively. {\bf (b)-}Saturated absorption spectrum for the transition $5{\rm S}_{1/2}, F_{g}=2\rightarrow 5{\rm P}_{3/2}$ of $^{87}\rm Rb$. Six Doppler-free dips are resolved corresponding from left to right to $F_{g}=2\rightarrow F_{e}=1$, crossover $F_{e}=1\&2$, $F_{e}=2$, crossover $F_{e}=1\&3$, crossover $F_{e}=2\&3$, and $F_{e}=3$. {\bf (c)-} Saturated absorption spectrum for the transition $5{\rm S}_{1/2}, F_{g}=3\rightarrow 5{\rm P}_{3/2}$ of $^{85}\rm Rb$. From left to right the peaks correspond to $F_{e}=2$, crossover $F_{e}=2\&3$, $F_{e}=3$, crossover $F_{e}=2\&4$, crossover $F_{e}=3\&4$, and $F_{e}=4$. Solid lines are data fitting using the product of six Lorentzian functions with a Gaussian profile. For the hyperfine structure of the $5{\rm P}_{3/2}$ excited state, we finally obtain
$\delta\left[(1\leftrightarrow 2)\right]_{^{87}Rb}= 156\pm 2 {\rm MHz}$,
$\delta\left[(2\leftrightarrow 3)\right]_{^{87}Rb}= 264\pm 3 {\rm MHz}$,
$\delta\left[(2\leftrightarrow 3)\right]_{^{85}Rb}= 64\pm 1 {\rm MHz}$, and
$\delta\left[(3\leftrightarrow 4)\right]_{^{85}Rb}= 119\pm 2 {\rm MHz}$, where $\delta\left[(i\leftrightarrow j)\right]$ is the energy splitting between hyperfine sublevels $F_{e}=i$ and $F_{e}=j$.
}
\label{Sat}
\end{figure}

The experimental setup used to implement Doppler-free saturated absorption spectroscopy of rubidium is described in figure~\ref{Sat}(a).\\
\indent The laser beam, tuned close to the rubidium transition $5{\rm S}_{1/2}\rightarrow 5{\rm P}_{3/2}$, is first sent through a half-wave plate and a polarizing beamsplitter (PBS), used to control the light power entering into the rubidium cell. The pump beam $P$ then travels through the rubidium cell and is retroreflected as a counterpropagating probe beam $S$ using a mirror and a $1\%$ optical density. A quarter wave plate is inserted in order to rotate by $90^{\circ}$ the probe beam polarization which is finally reflected by the PBS and detected using a photodiode.\\
\indent Typical saturated absorption spectra are depicted in figures~\ref{Sat}(b) and (c). Within a Gaussian envelope resulting from Doppler broadening, six dips appear, three of them being related to the hyperfine structure of the excited state $5{\rm P}_{3/2}$ and the three others corresponding to crossover resonances.\\
\indent The linewidths of the saturated absorption dips are found equal to $\Delta\nu=22\pm1$~MHz for $^{87}\rm Rb$ and $\Delta\nu=20\pm1$ MHz for $^{85}\rm Rb$. The lifetime $\tau$ of the $5{\rm P}_{3/2}$ excited state in rubidium is about $28$ ns. The natural linewidth predicted by the Heisenberg uncertainty principle is then $2\gamma=1/(2\pi \tau)=6$ MHz, much smaller than the one measured.\\
\indent As illustrated by equation~(\ref{Bloch}), the spectral linewidth $\Delta \nu$ of the absorption dips strongly depends on the pump beam intensity which saturates the atomic transition. As such saturation is at the heart of the method, it is hard to reach the natural linewidth using Doppler-free saturated absorption spectroscopy. A possible method would consist in measuring the linewidth of the dips for different pump beam intensities and to take the asymptotic value of the linewidth at null intensity. However, the obtained value would still be bigger than the natural linewidth because of collisions and inhomogeneous transit time broadening resulting from the finite time of interaction between the atoms and the laser light~\cite{Cahuzac}.\\
\indent In the early seventies, saturated absorption spectroscopy led to huge improvements in metrology experiments~\cite{CagnacHydro}. However, this method is sensible to recoil effect when an atom absorbs or emits a photon, which induces a splitting of the lines~\cite{Hall,Letokhov}. For metrology applications, saturated absorption spectroscopy has thus quickly been superseded by other methods, like Doppler-free two-photon absorption spectroscopy.

\section{Doppler-free two-photon absorption spectroscopy}

The possibility to use two photon absorption as Doppler-free spectroscopy method has been first proposed by L. S. Vasilenko, V. P. Chebotaev and A. V. Shishaev~\cite{Shishaev} whereas the first experimental demonstration has been simultaneously obtained in 1974 by F. Biraben, B. Cagnac and G. Grynberg~\cite{Biraben_PRL1974,Grynberg_PhD} in Paris and by M. D. Levenson and N. Bloembergen~\cite{Levenson} in Harvard .\\
\indent Nowadays, Doppler-free two-photon absorption spectroscopy is still the most powerful method for high precision measurements of fundamental constants like the Lamb shift or the Rydberg constant~\cite{Biraben_PRL1998}.

\subsection{Principle}
\label{TWOPrinciple}

\begin{figure}  [t]
\centerline{\includegraphics[width=10cm]{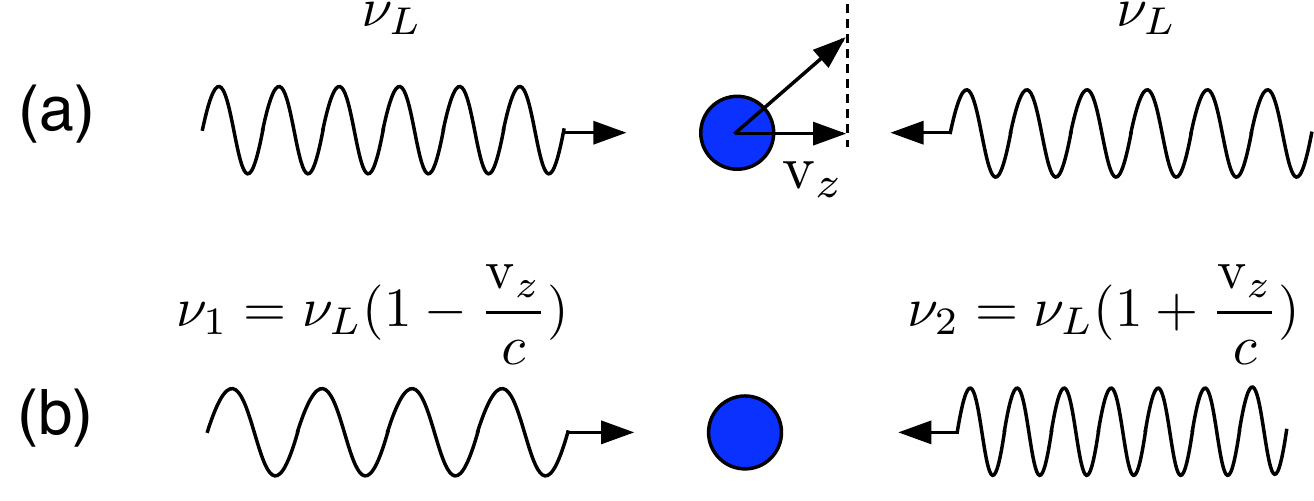}}
\caption{Principle of Doppler-free two-photon absorption spectroscopy. (a)-An atom is interacting with two opposite travelling waves at same frequency $\nu_{L}$ in the laboratory frame of reference. (b)-In the atom rest frame, the two frequencies are Doppler-shifted to $\nu_{1}$ and $\nu_{2}$. By absorbing two photons travelling in opposite directions, the Doppler shifts cancel.}
\label{Princ2P}
\end{figure}

\indent We consider atoms in a vapor interacting with two counterpropagating beams at same frequency $\nu_{L}$ in the laboratory frame of reference (figure~\ref{Princ2P}). In its rest frame, each atom with velocity ${\rm v}_{z}$ interacts with two Doppler shifted travelling waves at frequency $\nu_{1}$ and $\nu_{2}$ which, in the approximation of non-relativistic atoms ${\rm v}_{z}\ll c$, are equal to :
\begin{eqnarray}
\nu_{1}= & \nu_{L}\left[1-\frac{{\rm v}_{z}}{c}\right] \\
\label{deuxP1}
\nu_{2}= & \nu_{L}\left[1+\frac{{\rm v}_{z}}{c}\right] 
\label{deuxP2}
\end{eqnarray}
\indent We now suppose that the atoms can reach an excited state $|e\rangle$ (energy $E_{e}$) by absorbing two photons in the ground state $|g\rangle$ (energy $E_{g}$). Using equations~(7) and~(8), the resonant condition for absorbing two photons travelling in opposite directions is given by :
\begin{equation}
E_{e}-E_{g}=h\nu_{L}\left[1-\frac{{\rm v}_{z}}{c}\right]+h\nu_{L}\left[1+\frac{{\rm v}_{z}}{c}\right] = 2h\nu_{L}
\label{Reso}
\end{equation}
\indent The terms depending on atom velocity disappear as the two Doppler shift terms cancel. Consequently, when the resonance condition~(\ref{Reso}) is fulfilled, all the atoms, irrespective of their velocities, can absorb two photons. The line shape of this Doppler-free two-photon absorption resonance is theoretically predicted to be Lorentzian, the width being the natural linewidth $\gamma$ of the transition~\cite{Revue_Cagnac}. Contrary to saturated absorption spectroscopy, there is no recoil shift using such method as the total momentum transfer from light to atom is null.\\
\indent When the resonant condition is not realized, or when the two counterpropagating beams are not well superimposed, the atoms can not absorb two photons propagating in opposite directions anymore. However, some atoms can still absorb two photons propagating in the same direction, if their velocity ${\rm v}_{z}$ fulfill the relation $E_{e}-E_{g}=2h\nu_{L}\left[1\pm \frac{{\rm v}_{z}}{c}\right]$. Such a process leads to an additional Doppler-broadened profile. However, for a given frequency $\nu_{L}$ of the laser, only one class of atom velocities contributes to this Doppler broadened signal, whereas all the atoms contribute to the resonance signal when the condition~(\ref{Reso}) is fulfilled. As a result, Doppler-free two-photon absorption spectra appear as the superposition of a Lorentzian curve of large intensity and narrow width, and a Gaussian profile of small intensity and broad width (Doppler width)~\cite{Revue_Cagnac}. \\
\indent Single-photon transitions satisfy the selection rules $\Delta l=\pm 1$, where $l$ is the orbital angular momentum. Such transitions then require ground and excited states of opposite parity, like the transition $5{\rm S}_{1/2} \rightarrow 5{\rm P}_{3/2}$ investigated using saturated absorption spectroscopy in section~\ref{SatRes}. For two-photon transitions, the selection rules become $\Delta l=0, \pm 2$, allowing to investigate transitions between levels of identical parity. In that sense, two-photon spectroscopy is complementary to saturated absorption spectroscopy.\\
\indent In the following, we investigate the transition $5{\rm S}_{1/2} \rightarrow 5{\rm D}_{5/2}$ of rubidium. Two-photon absorption is achieved by exciting rubidium at the wavelength $\lambda=778.1$ nm (vacuum wavelength), and detected by monitoring the fluorescence at $420$ nm from the $5{\rm D}_{5/2} \rightarrow 6{\rm P}_{3/2}\rightarrow 5{\rm S}_{1/2}$ radiative cascade decay (Fig.~\ref{Setup_2P}(a)). \\
\indent Usually, non-linear processes like two-photon transitions require high-power lasers. However, for the case under study, the two-photon transition probability is greatly enhanced because the first excited $5{\rm P}_{3/2}$ state is close ($2$ nm detuning) to the virtual intermediate state involved in the two-photon process (Fig.~\ref{Setup_2P}(a)). Furthermore, the need of high power is also compensated by the fact that all the atoms contribute to the Doppler-free signal whereas saturation spectroscopy only involves a single class of atom with velocity ${\rm v}_{z}=0$. Therefore, the experiment can be performed with a simple tunable laser-diode~\cite{Olson,JOSA_Ryan,Ko_OptLett2004,Note}.  

\subsection{Experimental setup}   

\begin{figure}  [b]
\centerline{\includegraphics[width=10cm]{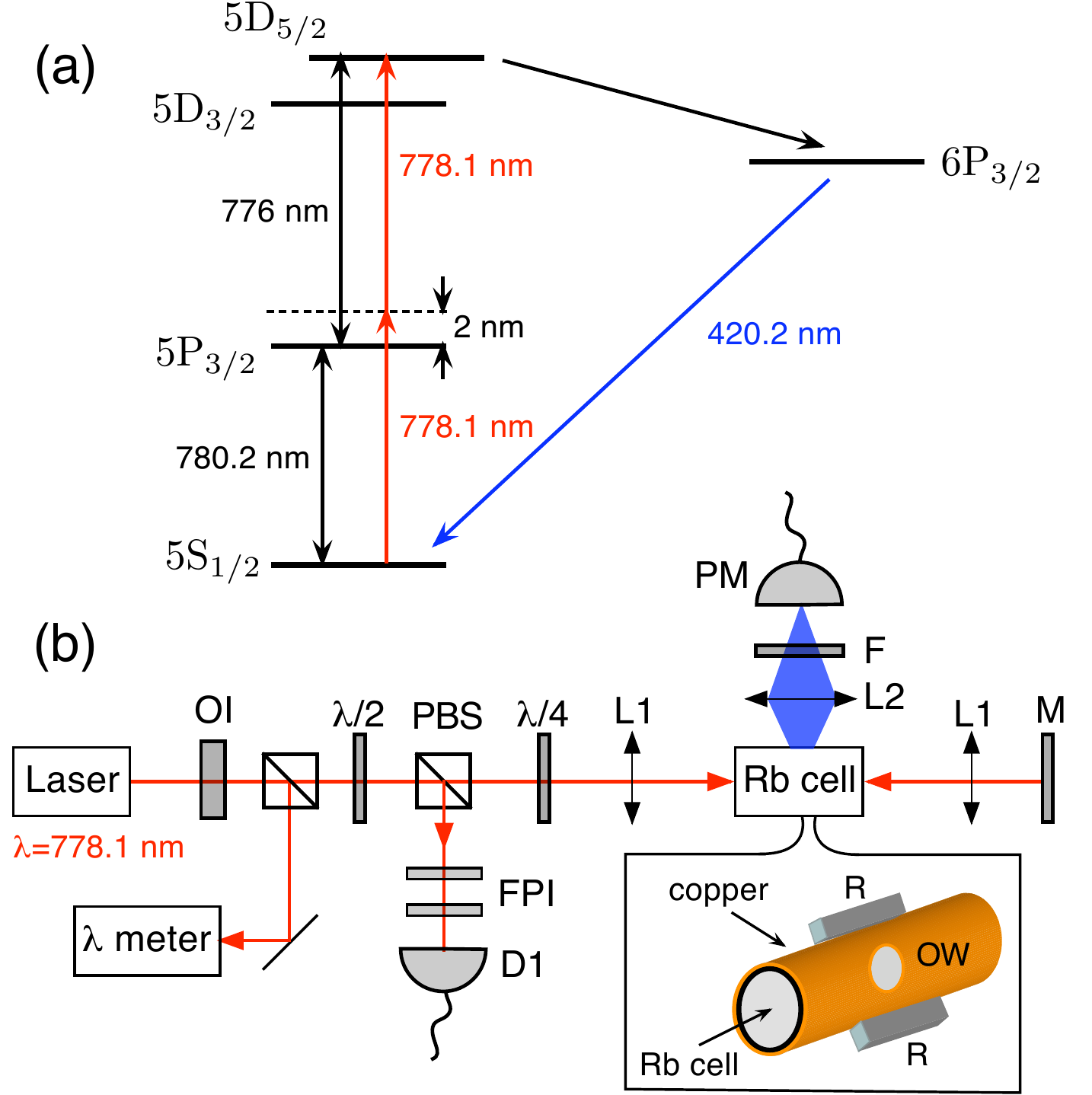}}
\caption{(a)-Energy levels involved in the two-photon transition $5{\rm S}_{1/2} \rightarrow 5{\rm D}_{5/2}$ of rubidium. (b)- Doppler-free two-photon spectroscopy setup. OI: optical isolator, $\lambda /2$: half-wave plate, PBS: polarization beamsplitter, $\lambda /4$: quarter-wave plate, FPI: Fabry-Perot Interferometer, D1 : photodiode, L1: $100$ mm focal lenses, M: mirror, L2: $60$ mm focal lens, F: $10$ nm width interference filter centered at $420$ nm. PM: photomultiplier. The rubidium cell is inserted in a copper ring, on which a couple of high power resistors (R) are glued in order to heat the cell. An optical window (OW) is cut inside the copper to detect the fluorescence.}
\label{Setup_2P}
\end{figure}

\indent The experimental setup used to implement Doppler-free two-photon absorption spectroscopy is depicted on figure~\ref{Setup_2P}(b).\\
\indent After preliminary tuning of the laser diode at the wavelength $\lambda=778.1$~nm, the laser beam is sent through a half-wave plate and a polarizing beamsplitter (PBS). The reflected light on PBS is directed into the Fabry-Perot Interferometer for frequency reference while the transmitted part is focused inside the rubidium cell using a pair of collimating lens. The beam is then retroreflected using a mirror to obtain the required counterpropagating beams configuration. For the present experiment, the laser power entering the rubidium cell is measured equal to $18$~mW. A quarter-wave plate is also introduced before the rubidium cell in order to excite the atoms with two counterpropagating beams with identical circular polarizations $\sigma^{+}$. This corresponds to the most efficient procedure to fulfill the selection rule $\Delta l=2$ of the investigated two-photon transition. Such configuration also strongly reduces the amount of light reflected into the laser diode, as the reflected light is mostly cut by the PBS. Even by using an optical isolator, we noticed that optical feedback was causing instability in the laser output when the quarter-wave plate was removed.\\
\indent The efficiency of two photon absorption strongly depends on the atomic vapor temperature~\cite{Olson}, which is directly related to the average number of atoms in interaction with the laser in the focal volume. For the studied two-photon transition, it has been shown that the fluorescence signal at $420$ nm used to monitor two-photon absorption begins to be efficient for temperature higher than $80^{\circ}$C~\cite{Olson}. This signal then increases with temperature until saturation around $130^{\circ}$C. At higher temperatures, even if the two-photon absorption probability is still increasing, the fluorescence signal did not increase anymore because of self-absorption of the $6{\rm P}_{3/2} \rightarrow 5{\rm S}_{1/2}$ transition~\cite{JOSA_Ryan}.\\
\indent Usually the atomic vapor is heated by introducing the cell inside an oven. Here the rubidium cell is simply inserted in a copper ring on which two high power resistors ($67 \ \Omega$, $10$ mW maximal power) are glued and used to heat the sample (see Fig.~\ref{Setup_2P}(b)). All contacts are realized with heat pasting grease and the ensemble is finally isolated using aluminium paper. An optical window is cut inside the copper ring in order to collect the blue fluorescence, using an imaging lens, an interferometric filter centered at the wavelength $\lambda=420$~nm, used to isolate the fluorescence from scattered light, and a photomultiplier.

\subsection{Results} 

\begin{figure}  [t]
\centerline{\includegraphics[width=11cm]{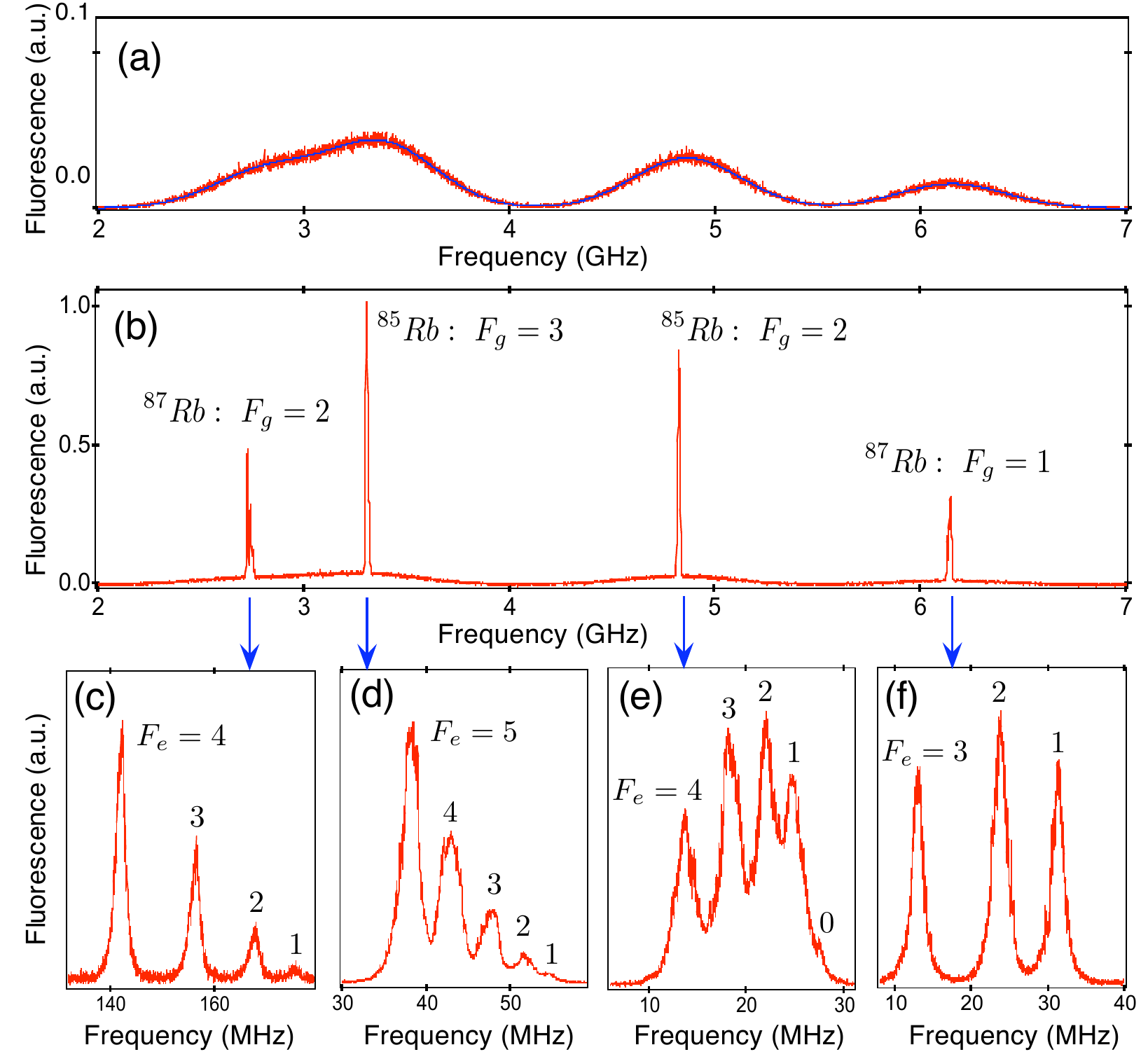}}
\caption{(a)-Fluorescence spectrum of the $5{\rm S}_{1/2} \rightarrow 5{\rm D}_{5/2}$ two-photon transition in rubidium, when the retroreflected beam is misaligned. Blue solid line correspond to a fit of the data with four Gaussian functions, which correspond to the ground state hyperfine levels of $^{85}\rm Rb$ and $^{87}\rm Rb$. Energy differences in term of laser frequency are related to hyperfine energy splittings by a factor one half. (b)-Doppler-free spectrum of the $5{\rm S}_{1/2} \rightarrow 5{\rm D}_{5/2}$ two-photon transition in rubidium. (c)-(f)-Hyperfine structure of the excited state $5{\rm D}_{5/2}$ for $^{85}\rm Rb$ and $^{87}\rm Rb$. $F_{g}$ and $F_{e}$ are respectively related to the total angular momentum of the hyperfine levels on the ground state and on the excited state. Data fitting with Lorentzian functions leads to $\delta\left[(1\leftrightarrow 2)\right]_{^{87}Rb}= 15 \pm 1 {\rm MHz}$, $\delta\left[(2\leftrightarrow 3)\right]_{^{87}Rb}= 22 \pm 1 {\rm MHz}$, $\delta\left[(3\leftrightarrow 4)\right]_{^{87}Rb}= 28 \pm 1 {\rm MHz}$, $\delta\left[(1\leftrightarrow 2)\right]_{^{85}Rb}= 5 \pm 1 {\rm MHz}$, $\delta\left[(2\leftrightarrow 3)\right]_{^{85}Rb}= 8 \pm 1 {\rm MHz}$, $\delta\left[(3\leftrightarrow 4)\right]_{^{85}Rb}= 10 \pm 1 {\rm MHz}$, and $\delta\left[(4\leftrightarrow 5)\right]_{^{85}Rb}= 9 \pm 1 {\rm MHz}$, where $\delta\left[(i\leftrightarrow j)\right]$ is the energy splitting between hyperfine sublevels $F_{e}=i$ and $F_{e}=j$.
}
\label{General_2P}
\end{figure}

Two-photon absorption spectra are recorded by scanning the frequency of the laser diode and detecting the fluorescence from the $5{\rm D}_{5/2} \rightarrow 6{\rm P}_{3/2}\rightarrow 5{\rm S}_{1/2}$ radiative cascade decay. When the retroreflected beam is misaligned, Doppler-broadened absorption spectra, which arise from absorption of two photons from the same laser beam, are measured. As depicted on figure~\ref{General_2P}(a), four lines corresponding to the hyperfine ground states of $^{85}\rm Rb$ and $^{87}\rm Rb$ are observed. However it is not possible to resolve excited state hyperfine structure, which remains hidden by Doppler broadening. By fitting each line with a Gaussian profile and using relation~(\ref{gaussien}), the temperature of the atomic vapor in the interaction volume can be estimated. A temperature of about $140 ^{\circ}$C is achieved, which corresponds to optimal detection of the two-photon absorption process, as discussed in the previous section.\\
\indent When the retroreflected beam is properly aligned, Doppler-free two-photon absorption spectrum is evidenced as depicted on figure~\ref{General_2P}(b). Strong fluorescence peaks at atomic resonant frequencies are superimposed to the remaining weak Doppler broadened profile. As a two-photon transition is considered, energy difference in terms of laser frequency is related to energy splitting of the atom by a factor one half. The ground state hyperfine splitting is then measured to be $3.03\pm 0.03$ GHz for $^{85}\rm Rb$ and $6.83\pm 0.06$ GHz for $^{87}\rm Rb$, in excellent agreement with other published values~\cite{Biraben_OptCom}. \\
\indent The hyperfine structure of the excited state $5{\rm D}_{5/2}$, is then studied by reducing the frequency scanning range of the laser diode and zooming on the resonances of the two-photon absorption profile. As depicted on figures~\ref{General_2P}(c)-(f), all the hyperfine levels of the $5{\rm D}_{5/2}$ excited state are properly resolved, except $F_{e}=0$ for $^{85}\rm Rb$. Even if the precision of the measurements is poor, due to laser diode instabilities, the measured spectral features are again in agreement with precise measurements given in reference~\cite{Biraben_OptCom}. Note that such an experiment also illustrates the selection rule on the total angular momentum $\Delta F=0,\pm1,\pm2$ for the two-photon absorption process

\begin{figure}  [b]
\centerline{\includegraphics[width=10cm]{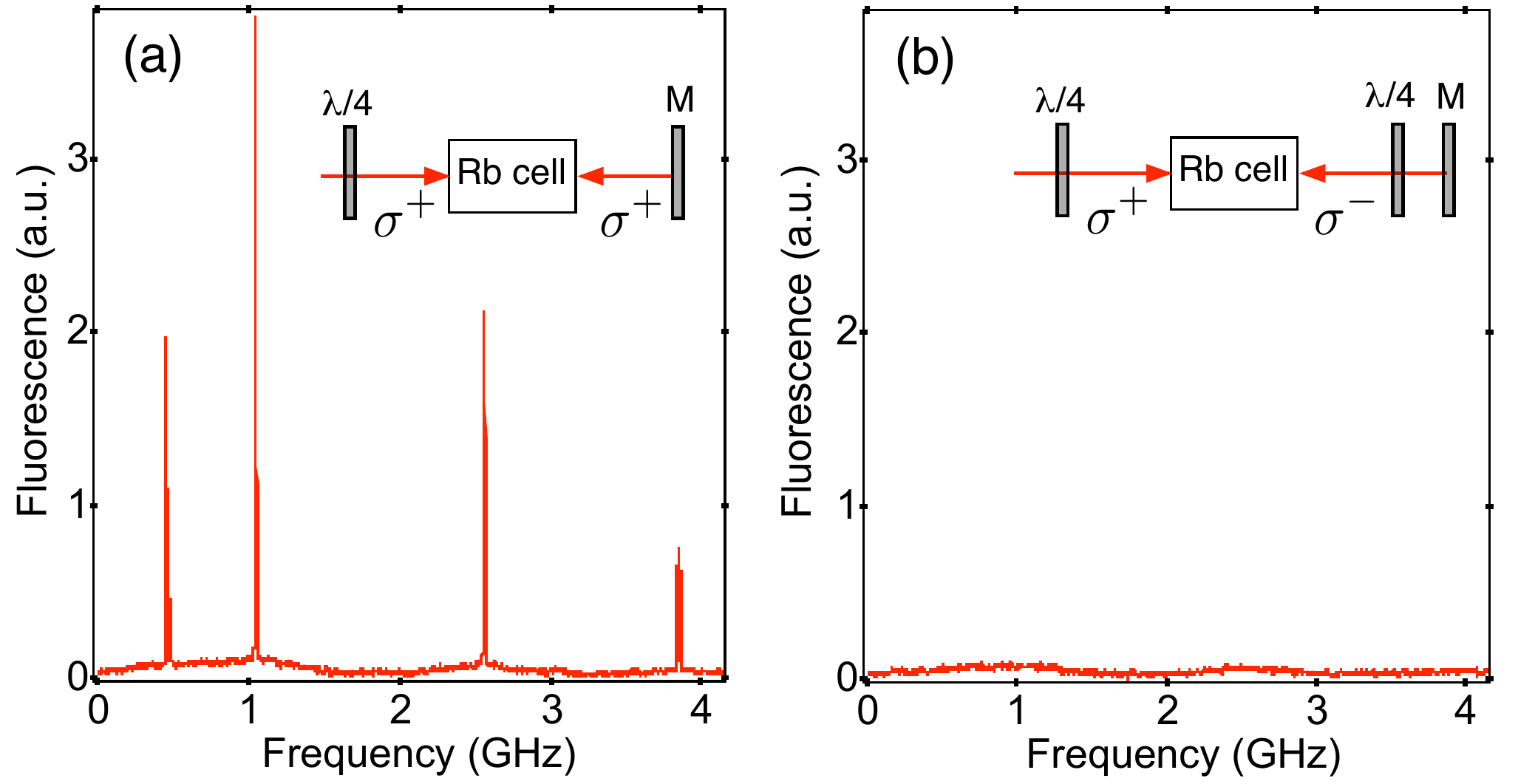}}
\caption{Illustration of two-photon transition selection rule $\Delta l=2$, by changing the beams polarizations. (a)-Fluorescence spectrum with two counterpropagating beams with identical polarization $\sigma^{+}$ and (b) with opposite circular polarization $\sigma^{+}$ and $\sigma^{-}$.}
\label{Selec}
\end{figure}

\indent For the data depicted on figure~\ref{General_2P}, the linewidths of the resonant peaks, measured in terms of laser frequency, are on the order of $2$ MHz, which is close to the spectral bandwith of the laser diode. The lifetime of the $5{\rm D}_{5/2}$ excited state is about $266$ ns, leading to a natural width of the transition around $2\gamma=600$ kHz. As two-photon transitions are considered, the measured width in terms of laser frequency is related to the transition linewidth by a factor one-half. To evidence the natural width of the transition, a laser with less than $300$ kHz spectral bandwidth would then be required.\\
\indent The experiment can be used to illustrate two-photon absorption selection rules by changing the laser beam polarizations. As mentioned before, the introduction of a quarter wave-plate before the rubidium cell allows to excite the atoms with two counterpropagating beams with identical circular polarizations $\sigma^{+}$. In that condition, the selection rule $\Delta l=2$ is always fulfilled by absorbing two photons and a strong Doppler-free signal is observed. Besides, if a second half-wave plate is introduced in front of the reflecting mirror, the polarization of the retroreflected beam becomes $\sigma^{-}$.  As a result, the absorption of two photons from opposite directions would lead to $\Delta l=0$, and no Doppler-free signal is observed, as depicted on figure~\ref{Selec}. However, the Doppler-broadened profile is still present because it corresponds to the absorption of two photons from the same beam ($\Delta l=2$). Note that the experimental configuration using two quarter-wave plate would be the optimal situation for the investigation of $S\rightarrow S$ transition ($\Delta l=0$), like the $5{\rm S}_{1/2} \rightarrow 7{\rm S}_{1/2}$ transition in rubidium. In that case, the profile of absorption spectra is just a Lorentzian curve, as the absorption of two photons from the same beam lead to $\Delta l=2$. The Doppler-broadened profile can then be eliminated for that particular case~\cite{Revue_Cagnac,Ko_OptLett2004}.

\begin{figure}  [h!]
\centerline{\includegraphics[width=10cm]{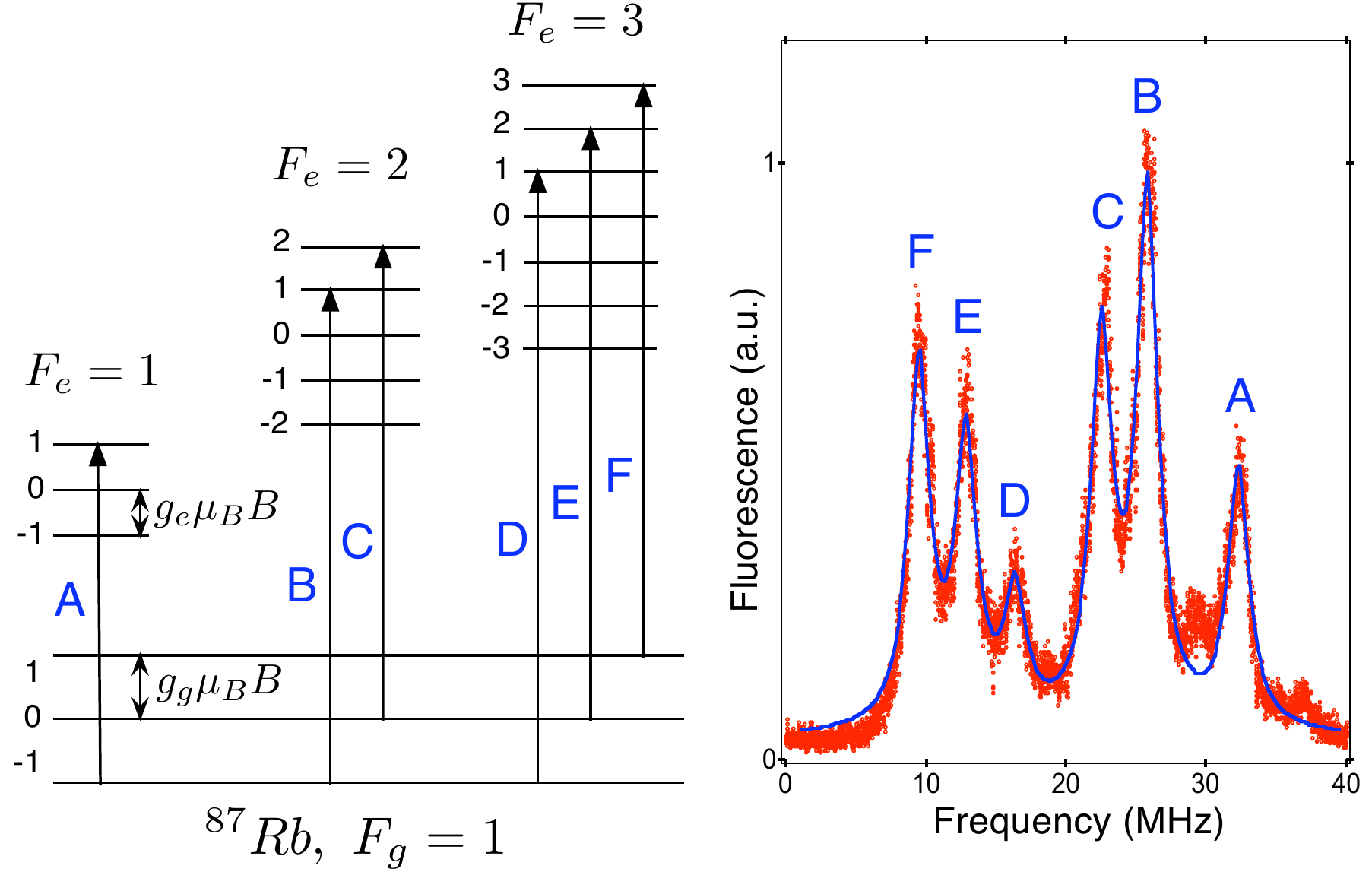}}
\caption{Zeeman splitting of the transitions between $5{\rm S}_{1/2},F_{g}=1$ and $5{\rm D}_{5/2}$ of $^{87}\rm Rb$, when a magnetic field $B$ is applied to the rubidium cell. A, B, C, D, E, F are related to the six transitions fulfilling the selection rule $\Delta m$=2.}
\label{Zeeman}
\end{figure}

\indent As a final experiment, a permanent magnet is put close to the cell in order to observe Zeeman splitting. A typical result is depicted on figure~\ref{Zeeman} for the transition $5{\rm D}_{1/2},F_{g}=1\leftrightarrow 5{\rm D}_{5/2}$ of $^{87}\rm Rb$. The transition through $F_{e}=1$ is not split, the one through $F_{e}=2$ is split into two lines and the one through $F_{e}=3$ is split into three lines. Such a result is an illustration of the transition selection rule $\Delta m=2$ (see Fig.~\ref{Zeeman}).\\
\indent In terms of energy, the splitting between consecutives Zeeman levels in the ground state is equal to $g_{g}\mu_{B}B$, where $g_{g}$ is the electron gyromagnetic factor in the ground state, $\mu_{B}$ the Bohr magneton and $B$ the magnetic field magnitude (see Fig.~\ref{Zeeman}). For the excited state, the Zeeman level splitting follows the same relation replacing $g_{g}$ by the electron gyromagnetic factor in the excited state $g_{e}$. As a result, it is straightforward to show that the frequency splitting of transitions associated with a given value of the excited state total angular momentum $F_{e}$ is given by $(g_{e}-g_{f})\mu_{B}B$.\\
\indent A systematic study of  Zeeman splitting as a function the magnetic field magnitude has not been performed yet. As a next step, we plan to introduce the rubidium cell inside coils in order to have better control of the applied magnetic field. This configuration would allow to measure the parameter $(g_{e}-g_{f})$  or to use the Doppler-free two photon absorption spectrum as an atomic magnetometer.

\section{Conclusion}

While atomic physics may be considered by students as a theoretical subject, we present here experiments enabling advanced undergraduates to become acquainted with spectroscopy techniques widely used in laboratories. Throughout this article, we have implemented two complementary methods of high resolution spectroscopy - saturated absorption and two-photon spectroscopy- to probe the hyperfine structure of Rubidium.
Eventually, these experiments combined within a single setup, constitute a great opportunity for students to investigate atomic spectra and to illustrate quantum mechanics selection rule as well as Zeeman Effect.\\

\ack{The authors are grateful to F.~Biraben for fruitful discussions and C. Ollier for technical support. We acknowledge financial support by Ecole Normale Sup\'eriere de Cachan.}


\end{document}